\documentclass[aps,prd,superscriptaddress,twocolumn,,amssymb,eqsecnum,showpacs,showkeyes,secnumarabic,graphics,floatfix,nofootinbib,tightenlines,longbibliography]{revtex4-1}
\usepackage{float}
\usepackage{graphicx}
\usepackage{epsfig}
\usepackage{bm}
\usepackage{cancel}
\usepackage{amsmath}
\usepackage[toc,page]{appendix}
\usepackage[dvipsnames]{xcolor}
\usepackage[margin=3cm]{geometry}
\usepackage{hhline}
\usepackage[utf8]{inputenc}
\usepackage[breaklinks=true,colorlinks=true,
linkcolor=blue,urlcolor=blue,citecolor=cyan,
bookmarks=true,bookmarksopenlevel=2]{hyperref}

\def \beq  {\begin{equation}}
\def \eeq  {\end{equation}}
\def \ber  {\begin{eqnarray}}
\def \eer  {\end{eqnarray}}

\begin{document}
\newcommand{\newc}{\newcommand}

\newc{\be}{\begin{equation}}
\newc{\ee}{\end{equation}}
\newc{\ba}{\begin{eqnarray}}
\newc{\ea}{\end{eqnarray}}
\newc{\bea}{\begin{eqnarray*}}
\newc{\eea}{\end{eqnarray*}}
\newc{\ie}{{\it i.e.} }
\newc{\eg}{{\it e.g.} }
\newc{\etc}{{\it etc.} }
\newc{\etal}{{\it et al.}}
\newc{\lcdm}{$\Lambda$CDM}
\newcommand{\nn}{\nonumber}

\date{\today}
\title{\bf  Black hole or Gravastar? The GW190521 case}

\author{Ioannis Antoniou}\email{i.antoniou@uoi.gr}
\affiliation{Department of Physics, University of Ioannina,
GR-45110, Ioannina, Greece}

\begin{abstract}
The existence of cosmological compact objects with very strong gravity is a prediction of General Relativity and an exact solution of the Einstein equations. These objects are called black holes and recently we had the first observations of them. However, the theory of black hole formation has some disadvantages. In order to avoid these, some scientists suggest the existence of gravastars (gravitation vacuum stars), an alternative stellar model which seems to solve the problems of the black hole theory. In this work we compare black holes and gravastars using a wide range of the literature and we emphasize the properties of gravastars, which are consistent with the current cosmological observations. Also, we propose gravastars as the solution of the "pair-instability" effect and a possible explanation for the observed masses of the compact objects, before the collapse, from the gravitational signal GW190521, since in the formation of a gravastar there aren't mass restrictions.
\end{abstract}
\maketitle

\begin{section}{Introduction}
One of the most attractive concepts in General Relativity is the existence and the properties of black holes, a region of spacetime where gravity is so strong that nothing, no particles or even electromagnetic radiation such as light, can escape from it. A black hole is characterized by the charge $Q$, the mass $M$ and the  total angular momentum $J$. The mass $M$ is in the range above $3M_{\bigodot}$ until a few decades of the mass of Sun, or a few millions or billions $M_{\bigodot}$. The latter are known as supermassive black holes and exist in the centers of most galaxies. Even if the theory of black hole formation is well known, there are some abstruse and 'strange properties'  such as
\begin{itemize}
    \item the "event horizon", a boundary in spacetime through which matter and light can pass only inward towards the mass of the black hole.
    \item the gravitational singularity at the center of a black hole, a region where the spacetime curvature becomes infinite.
    \item the information paradox, where a pure quantum state which passes over the event horizon can evolve into a mixed state during the evaporation of the black hole.
\end{itemize}
The first picture of the structure in the center of M87 \cite{Akiyama:2019cqa} is a strong evidence for the existence of black holes, or some other compact object with very strong gravity. Many scientists dispute the existence of black holes because if we take into account quantum effects, the gravitational collapse of objects comes to a halt and furthermore no event horizon forms \cite{Frolov:2014wja}.  Cosmologists all around the world investigate the existence of an alternative model \cite{Cardoso:2019rvt}, which has not the above 'strange properties' \cite{Cattoen:2006tk}. Consequently, this model will be more attractive and realistic than black holes \cite{Konoplya:2019nzp, Dymnikova:2015yma}. The most successful cosmological model in this direction is called gravastar configuration, an alternative endpoint of gravitational collapse of a massive star, which do not involve horizon and proposed by  Mazur and Mottola \cite{Mazur:2001fv}. The main idea is that the gravitational collapse stops at a radius greater than the radius of the horizon \cite{Nakao:2018knn}. Mazur and
Mottola also showed that their gravastars are thermodynamically stable unlike in the case of the black hole.
\end{section}

\begin{section}{The structure of a Gravastar}
Thin spherical shells are boundary
hypersurfaces with shell radius $R$, surface energy density $\sigma$ and surface
pressure $p$. Gravastar is a spherical static stable thin shell configuration and an exact solution of the equations of motion in GR when the shell is spherical and infinitesimally thin \cite{Israel:1966rt, Hoye:1983dq, Misner:1974qy}. Usually a gravastar is divided in three regions. The interior geometry is de Sitter metric with equation of state $p=-\rho$ (false vacuum or dark energy) and the exterior is usually Schwarzschild metric with equation of state $p=0=\rho$ (true vacuum). The exterior metric is related to the interior metric in the context of the Israel junction conditions \cite{Israel:1966rt}.  On the horizon there is a thin shell of matter (perfect fluid) with radius $R$ and equation of state $p=\rho$.

The concept of gravastar is related to a Riemannian manifold which is divided by a 3-dimentional timelike surface in two pieces, named exterior and interior regions and are gluing on this surface. Visser \cite{Visser:2003ge} introduced such an approach to develop thin-shell gravastars by the junction of both spacetimes that eliminate the event horizon and singularity. The mathematical procedure used to develop the geometry of the gravastars is based (mainly) on Israel's \cite{Israel:1966rt} and Mazur and Mottola works \cite{Mazur:2001fv}. The metric has the form 
\begin{equation}\label{e1}
ds_{\pm}^2=-f_{\pm}(r)dt_{\pm}^2+\frac{dr_{\pm}^2}{f_{\pm}(r)}+r_{\pm}^2(d\theta_{\pm}^2+\sin^2{\theta_{\pm}}d\phi_{\pm}^2)
\end{equation}
where the (+) corresponds to the exterior region, while the (-) to interior region. The $f_\pm (r)$ function determines the interior and exterior background of the thin shell. 

The crucial question about the concept of gravastars is under which conditions these configurations are stable. Many properties affect the answer, such as the background geometry \cite{Alestas:2020wwa}, radial perturbations (using static spherical models) about the equilibrium shell's radius \cite{Visser:2003ge, Lobo:2020kxn, Sharif:2020bzt}, the function of the innermost and exterior mass \cite{Visser:2003ge} and the rotation of the shell \cite{Cardoso:2007az, Chirenti:2008pf}. In every case there are values of the corresponding parameters for which the potential $V(r)$ of the shell has minimum equal to zero. Actually, there is a wide range of parameters which allow stable gravastar solutions \cite{Carter:2005pi, Alestas:2020wwa}, which means that the gravastar is a viable cosmological structure. 

In order to construct a stable gravastar, many authors consider a shell with different cosmological backgrounds (except the Schwarzshild-de Sitter) such as a background with cosmological constant \cite{Yamanaka:1992zd} or regular spacetimes (Bardeen and Bardeen-de Sitter black holes) \cite{Sharif:2020bzt}. They investigate the stability under radial perturbations \cite{Visser:2003ge, Alestas:2020wwa} or under slow rotation \cite{Uchikata:2015yma, Chirenti:2008pf, Cardoso:2007az}. Rotating gravastars can affect the inertial frames and they induce the dragging effect \cite{Cohen:1966kg, Lindblom:1974bq}. The rotation of a massive shell induces rotation of the inertial frame which tend to be equal near the Schwarzschild radius and the results of the rotation and expanding/re-collapsing shell are examined for their consistency with particular interpretations of Mach's principle.

The detection of gravitational waves has opened a new window in the Universe and a new way to observe cosmological structures and signatures of them. Many authors have investigated the behaviour of a gravitation wave in the vicinity of a compact mass. Authors of Ref. \cite{Bishop:2019ckc} found that the dust shell causes the gravitational wave to be modified both in magnitude and phase without energy transfer, while the authors of Ref. \cite{Antoniou:2016vxg} found that a compact mass such as a black hole induces modification in frequency, magnitude and energy of a wave in the vicinity of the mass. The different behaviour  could be a criterion to distinguish a gravastar from a black hole.

Another way to discriminate a gravastar from a black hole is developing in Ref. \cite{Pani:2009ss}. Quasinormal frequency modes (complex numbers where the imaginary part corresponds to the loss of energy) are produced from a black hole. The authors of Ref. \cite{Pani:2009ss} computed polar and axial oscillation modes of gravastar. They found that  the quasinormal mode spectrum is completely different from that of a black hole when both have the same gravitational mass. Also, the equation of state of the matter-shell affects the polar spectrum. In Ref. \cite{Chirenti:2007mk} the authors calculated quasinormal modes of axial parity perturbations and they found that the decay rate of a black hole and a gravastar are not the same.

The evolution of a star with mass above $3M_{\bigodot}$ ends with a black hole, but there is an upper mass-limit. The mass must be less than $64M_{\bigodot}$ \cite{Woosley:2016hmi}. From stellar evolution in close binaries, no black holes between  $52M_{\bigodot}$ and $133M_{\bigodot}$ are expected. The effect is known as “pair-instability” \cite{Leung:2019fgj}. The GW190521 shows that there are compact objects with mass in the above range (confidence level $99\% $) and the scientific community is looking for answers \cite{Abbott:2020mjq}. The mass of the initial compact objects was $85^{+21}_{-14}M_{\bigodot}$ and $66^{+171}_{-18}M_{\bigodot}$ ($90\%$ credible intervals), which falls in the mass gap predicted by pulsational pair-instability supernova theory \cite{Abbott:2020tfl}. The mass of the remnant after the collapse is $142^{+28}_{-16}M_{\bigodot}$ which is near the above limit of the mass gap.

In order to calculate the total mass of a gravastar, we must divide it in three regions, the inner region, the thin shell and the outer region. In the case where the interior region has negative pressure with equation of state $p=-\rho$ (de Sitter) the inner mass is \cite{Abbas:2020kju} 
\begin{equation}\label{e2}
    m_{-}(R)=\int_0^R 4\pi r^2 c_0 dr=\frac{4}{3}\pi R^3c_0
\end{equation}
where $c_0$ is the constant matter density throughout the interior region. The parameter $c_0$ includes information about the constant pressure and constant density throughout the interior region. 

The mass of the ultrarelativistic, extremely thin stiff shell  \cite{Debnath:2019eor} of radius $R$, which obeys the equation of state $p=\rho$ (the density is very high) and has width $\epsilon\ll1$ is 
\begin{equation}\label{e3}
    m_s=4\pi R^2 \sigma
\end{equation}
where $\sigma$ is the surface energy density of the thin shell. 

The exterior region is a static Schwarzschild geometry with equation of state $p=\rho=0$ and the parameter $m_{+}$ in the metric is the total mass of the stellar structure of a gravastar. Thus, the total mass \cite{Debnath:2019eor} is the sum of the equations \eqref{e2} and \eqref{e3}, ie
\begin{equation}\label{e4}
    M_{gr}\equiv m_{+}=m_{-}+m_s=4\pi R^2(c_0\frac{R}{3}+\sigma)
\end{equation}
This is the mass which is observable from the Earth for this stellar object. From eq. \eqref{e4} we conclude that the mass $M_{gr}$ depends on the radius of the thin shell, the matter density of the inner region and the surface energy density of the thin shell. The result could be as big as needed, to support the cosmological observations.

\begin{figure}[ht!]
\centering
\includegraphics[width = 0.48
\textwidth]{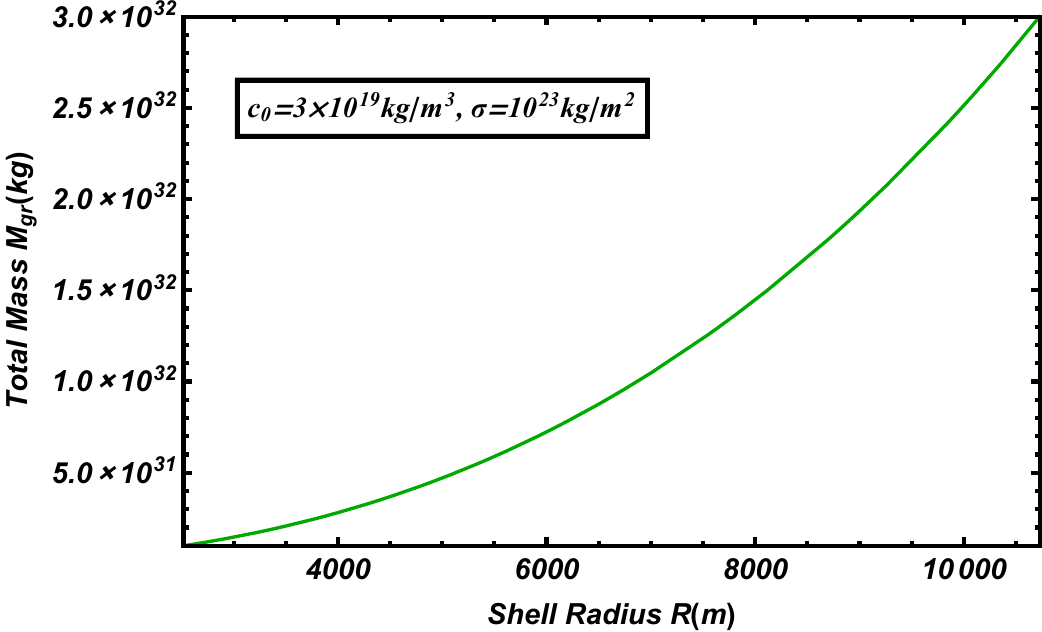}
\caption{The total mass of the gravastar in the range ($10M_\odot, 150M_\odot$) as a function of its radius, when the matter density and surface energy density are stable.}
\label{fig1a}
\end{figure} 

For example, if the mass of the gravastar is $M=10^3M_\odot$ the corresponding Schwarzschild radius is $R_s=3\times 10^3km$. We consider the radius of the gravastar as $R=10^4km$ and we conclude that $c_0\frac{R}{3}+\sigma\simeq 10^{23}kg/m^2$. Thus, the order of the surface energy density of the thin shell must be $\sim 10^{23}kg/m^2$ and the matter density of the inner region $\sim 10^{19}kg/m^3$. 

In figure \ref{fig1a} we have plot the mass of the gravastar (in the pair instability region of a black hole) as a function of its radius, when the matter density is $c_0=3\times 10^{19}kg/m^3$ and the surface energy density $\sigma=10^{23}kg/m^2$. 

It is obvious that the theory of gravastars hasn't mass restrictions, so it seems that this configuration can solve the problem of the mass gap that GW190521 suffers from. This mass gap arises from the pair instability or pulsational pair instability causing mass loss or destruction of the stellar progenitor prior to the formation of any remnant. In the process of stellar evolution there could appear gravastars of masses from GW190521. Probably, this gravitational wave signal  (GW190521) is a strong indication for the existence of gravastars.

\end{section}

\begin{section}{The Hierarchical Scenario}
GW190521 is the most massive gravitational wave event observed to date. It was reported as a binary black hole merger, in which the inferred masses of the black holes in the binary place them (and the merger remnant) in the pair instability mass gap. A hierarchical scenario \cite{Gerosa:2021mno,Tagawa:2020qll} for the formation of the binary suffices for reconciling the observed masses with the pair instability mass gap, although it raises interesting questions about the environment in which such mergers would happen. 

Our purpose in this work is to suggest an alternative and possibly more realistic scenario than the hierarchical one, for the origin of the GW190521 and not to review the literature about possible explanations. In Ref. \cite{Gerosa:2021mno} and references therein, one can find a wide review. In this review \cite{Gerosa:2021mno}, especially in subsection 4.4, for the interpretation of the GW190521 through the hierarchic scenario are required 2 conditions
\begin{itemize}
    \item 4 black holes in the same region. From that we know until now, we haven't any observation or indication for this event.
    \item a special scenario for the collapse. In the first generation the black holes have merged two by two and in the second generation the new black holes have merged again. This scenario is called 2g+2g \cite{Tagawa:2020qll} (higher-g)!!
\end{itemize}
It is obvious that these conditions restrict the possibility of this scenario and we think that is less possible than our proposal (one way of possibly circumventing the limitations from the mass gap is by having a gravastar merger).

In Ref. \cite{Gerosa:2021mno}, the authors review several alternative explanations for the occurrence of
GW190521 such as population III stars at very low-metallicity,
accretion onto either stellar-origin or primordial
BHs, and stellar mergers. Another speculations include exotic compact objects (beyond $\Lambda$CDM model) and dark-matter annihilation. Also, there is possible
that the primary and secondary components of GW190521 are not inside the pair instability gap respectively, but above and below this gap. From all these, it is obvious that the GW190521 is an open issue in cosmology. 

In the literature \cite{Chirenti:2016hzd, Cardoso:2016rao}, there are previous works which connect the gravitational waves (mainly the first signal GW150914) with the existence of gravastars. The mass of the origin of these signals is outside the range of the mass gap (pair instability). Also, LIGO's observations of gravitational waves from colliding objects have been found to be indistinguishable from ordinary black holes \cite{Cardoso:2016rao}.\\

\end{section}

\begin{section}{Conclusions} 

Gravastar is a suggested (hypothetical) stellar compact object alternative to a black hole. It is a kind of spherical shells, which is a stable configuration under rotation and radial perturbations \cite{Antoniou:2022vxi}. Until now, there is no observational way to distinguish a black hole from a gravastar. Thus, a possible observation of a black hole can be a possible gravastar. Recent observations of gravitational wave GW190521 pointed out that the theory of black holes formation is not consistent with some observations, since there is a mass gap in their formation. From theoretical view, there is a way to avoid the pair instability effect/mass gap with gravastar formation. The GW190521 event could be an indication for the existence of gravastars, or spherical shells which might avoid the mass gap. This stellar configuration  (gravastar) is a promising candidate to explain the observations and to complete the puzzle. 
\end{section}

\section{Acknowledgements}
I would like to thank Professors Leandros Perivolaropoulos (University of Ioannina) and Demetrios Papadopoulos (Aristotle University) for useful  discussions and indications about the gravastars, which improved the quality of this work.

\raggedleft
\bibliography{gravastar}

\end{document}